\documentclass[twocolumn,floatfix, showpacs]{revtex4}

\usepackage[final]{epsfig}
\usepackage{amsmath}
\begin{document}
 
\title{A study of the constraining power of high $P_T$ observables in heavy-ion collisions }
 
\author{Thorsten Renk}
\email{thorsten.i.renk@jyu.fi}
\affiliation{Department of Physics, P.O. Box 35, FI-40014 University of Jyv\"askyl\"a, Finland}
\affiliation{Helsinki Institute of Physics, P.O. Box 64, FI-00014 University of Helsinki, Finland}


\begin{abstract}
Since the start of the LHC heavy ion program, a multitude of rather different high transverse momentum ($P_T$) observables has become available to study the physics of the interaction of hard partons with a QCD medium. Similarly, multiple  theoretical models for this interaction exist and have been compared with available data, and regularly physics conclusions are drawn based on the agreement of a model with a particular data subset. However, such an agreement is only a necessary condition to identify a physics mechanism, not a sufficient one --- it needs to be demonstrated that the agreement is not accidential or generic, in other words the observable actually needs to measure the physics in question. The aim of the study presented here is to illustrate this problem by computing various high $P_T$ observables in three different models, two of which are known to be grossly wrong \emph{ab initio}, and to study in which observables what amount of discrepancy with the data appears. The surprising outcome is that jet yield observables are not very sensitive to the physics mechanism of jet-medium interaction and hence do not constrain models well.
\end{abstract}
 
\maketitle

\section{Introduction}

While high $P_T$ observables in the context of A-A collisions at RHIC were comparatively scarce, the start of the LHC heavy-ion program has added a multitude of novel observables. Broadly, observables can now be classified into the following categories: 1) hadron-based observales, including the nuclear modification factor $R_{AA}^h$ measuring the disappearance of hard hadrons in the spectrum \cite{PHENIX-classic,PHENIX-new,ALICE-RAA,CMS-RAA} and the back-to-back hadron coincidence suppression $I_{AA}$ \cite{STAR-DzT,ALICE-h-h} 2) reconstructed jet-based observables, including the jet nuclear modification factor $R_{AA}^{jet}$ (or $R_{CP}^{jet}$) \cite{ALICE-jet-RAA,ATLAS-jet-RCP,ALICE-jet-RCP} and the modification of back-to-back jet coincidences \cite{CMS-jet-jet, ATLAS-jet-jet}  3) jet-triggered coincidence measurements in which a hadron recoiling from the jet is observed, such as back-to-back jet-hadron coincidences \cite{STAR-jet-h} and jet fragmentation measurements in which the distribution of hadrons inside a found jet is studied \cite{CMS-FF,ATLAS-FF} and 4) other (usually triggered) measurements such as photon-hadron \cite{PHENIX-gamma-h} or photon-jet \cite{CMS-gamma-jet} coincidences. From a more general point of view, all these measurements can be seen as observing the same physics of jet-medium interaction through the filter of different, observation-induced biases \cite{Bayesian}.

On the theory side, many different models have been proposed for the description of the physics of parton-medium interaction. Note that conceptionally not all of them are able to compute the whole set of available observables: For instance leading parton energy loss models (e.g. \cite{QuenchingWeights,ASW-1,HT-DGLAP,AMY-1,AMY-2,WHDG}) neglect the in-medium virtuality evolution of the jet and hence can be applied to hadronic observables only (for a more detailed discussion of the leading parton energy loss approximation, see e.g. \cite{Constraining}). On the other hand, several in-medium shower evolution codes do not include a hadronization stage yet (e.g. \cite{MATTER,XNW,Hybrid}) and can hence only be compared with jet observables. Comparatively few available models involve both an in-medium virtuality evolution and hadronization and can hence conceptually be applied to the full set of available observables \cite{JEWEL,YaJEM1,YaJEM2,Q-PYTHIA,MARTINI}.

The implication is that in many cases, model to data comparisons are not done for the full range of available high $P_T$ observables but only for  subset. An relevant question in this context is to what degree a model can still be constrained by such a subset of the data. Similarly, in order to argue that the data supports a certain physics assumption made in a model, demonstrating agreement between model and data is a necessary condition, but not a sufficient one. It may for instance be the case that a particular observable is driven by a combination of biases and known vacuum physics which makes the outcome generic and independent of any specific physics assumption for the effect of the medium, or the agreement could be due to an accidential cancellation of effects. To avoid these pitfalls, model scenarios which do not incorporate the physics assumption in question need to be tested and demonstrated to fail with the data.

The aim of this work is to present a sensitivity analysis of jet, hadron and mixed jet-hadron coincidence observables in this spirit. The basic idea is to compute the same set of observables within three different scenarios for parton-medium interaction. Each of the scenarios is chosen to represent a substantially different physics picture of jet quenching, and the aim of the study is to illustrate which observables are  suited to probe these differences by creating a large tension between data and model for wrong assumptions. This in turn establishes the idea to define a set of key observables which are particulary powerfu to constrain models, i.e. if models are compared to a subset of data only, it should be the subset of key observables for getting maximal constraints from the data.

\section{Parton-medium interaction scenarios}

All the different scenarios of parton-medium interaction to be tested in this study are available as modes of the in-medium shower evolution code YaJEM and documented in \cite{YaJEM2,YaJEM3} to which the reader is refered for more details. YaJEM is based on the PYSHOW algorithm \cite{PYSHOW} to which it reduces in the absence of medium effects, followed by a hadronization using the Lund model \cite{Lund}. 

All scenarios are embedded into a full fluid-dynamical model of the medium evolution, providing the local energy density $\epsilon(\zeta)$ as well as the flow  velocity $u^\mu(\zeta)$ at the spacetime position $\zeta$ of a parton probing the medium. In the case of LHC observables, the fluid dynamics is documented in \cite{hydro-LHC}, for RHIC kinematics in \cite{hydro3d}. Hard events are generated in momentum space according to leading order pQCD expressions supplemented with an intrinsic $k_T$ imbalance with a Gaussian distribution of 2 GeV width to mimic higher order effects, and  localized in the transverse plane according to the probability density for binary collisions.

The various scenarios for parton-medium interaction arise by characteristic modifications of the QCD shower evolution equations as the shower propagates through the medium. As discussed in \cite{YaJEM2,YaJEM3}, there are three different possibilities available within YaJEM.

The action of the medium can be parametrized by transport coefficients $\hat{q}(\zeta), \hat{e}(\zeta)$ which parametrize the virtuality transfer per unit pathlength and the energy loss per unit pathlength respectively. These coefficients are taken to be proportional to $\epsilon^{3/4}(\zeta) F(\rho(\zeta), \alpha(\zeta)$ with

\begin{equation}
\label{E-flow}
F(\rho(\zeta), \alpha(\zeta)) = \cosh \rho(\zeta) - \sinh \rho(\zeta) \cos\alpha(\zeta).
\end{equation}

a hydrodynamical flow correction factor accounting for the Lorentz contraction of the density of scattering centers as seen by the hard parton for $\rho(\zeta)$ the local flow rapidity and $\alpha(\zeta)$ the angle between hydrodynamical flow and parton propagation direction.

For an intermediate shower parton $a$, created at a time $\tau_a^0$ and existing for a duration $\tau_a$ before branching into a pair of daughter partons, the virtuality as propagated inside the shower code is then changed by 

\begin{equation}
\label{E-Qgain}
\Delta Q_a^2 = \int_{\tau_a^0}^{\tau_a^0 + \tau_a} d\zeta \hat{q}(\zeta)
\end{equation}

which opens the phase space for the possibility of additional, medium induced radiation. Similarly, the action of the transport coefficient $\hat{e}$ is to reduce the parton energy by

\begin{equation}
\label{E-Drag}
\Delta E_a = \int_{\tau_a^0}^{\tau_a^0 + \tau_a} d\zeta \hat{e}(\zeta)
\end{equation}

If the parton is a gluon, the energy and momentum transfer are increased by the ratio 2.25 of the gluon to quark Casimir color factors.

The scenario {\bfseries YaJEM-DE} \cite{YDE} is the current default of YaJEM and constrained by a large body of data. It utilizes both $\hat{q}$ and $\hat{e}$ with the relative proportion constrained by the data of back-to-back hadron correlations at RHIC \cite{STAR-DzT}, leading to about 10\% of the total energy loss from a leading parton driven by $\hat{e}$, and the majority of energy loss coming through the copious emission of soft gluons explicitly treated in the shower evolution. Since, as decribed above, $\hat{q}$ and $\hat{e}$ lead to corrections to parton kinematics at scales $\Delta Q^2, \Delta E$ which are properties of the medium and do not depend on the kinematics of the shower, the self-similarity of the fragmentation function is broken at a characteristic scale around 3 GeV \cite{jet-h}. At this scale, showers are also strongly broadened by the medium effect \cite{jet-h}.

As discussed in more detail in \cite{YD}, the in-medium part of the shower can persist only for a length $L$ before the shower emerges from the medium, this implies a minimum virtuality scale $Q_{min} \sim \sqrt{E/L}$ which also depends on the original parton energy $E$ down to which the in-medium shower can evolve. This in turn leads to a strong energy dependent and non-linear pathlength dependence of the medium modification, in agreement with data \cite{YD,RAA-LHC}.

The scenario {\bfseries YaJEM-E} utilizes only the transport coefficient $\hat{e}$ and hence does not lead to any additional medium induced radiation, all energy lost from the shower is assumed to be dissipated into the medium. As a result, there is no additional soft gluon radiation leading to copious production of soft hadrons, and showers become more collimated even at high $P_T$ \cite{YaJEM3}. Just as YaJEM-DE described above, it breaks the self-similarity of the fragmentation function at a fixed scale. The minimum virtuality scale down to which the shower is evolved in the medium is kept fixed at $Q_{min} = 1$ GeV, resulting in an approximately linear pathlength dependence of the medium modification and the leading parton energy loss.

Finally, the scenario {\bfseries YaJEM+BW} does not include any explicit energy-momentum transfer between shower and medium, but utilized the Borghini-Wiedemann (BW) prescription \cite{BW} for modifying the pQCD splitting kernels generating the fragmentation function to mimic medium effects.
In this scenario, the singular part of the branching kernel in the medium is enhanced by a factor $1+f_{med}$, e.g. the branching kernel for $q \rightarrow qg$  becomes in the medium

\begin{equation}
P_{q\rightarrow qg}(z) = \frac{4}{3} \frac{1+z^2}{1-z} \Rightarrow \frac{4}{3} \left( \frac{2 (1+f_{med})}{1-z} - (1+z)\right)
\end{equation}

where $z$ is the splitting variable for the parent parton energy among the two daughters produced in the branching.

This increased branching probability leads to additional, medium induced soft gluon production and widens the shower somewhat in transverse space, although no explicit flow of energy and momentum between jet and medium is modeled. The factor $f_{med}$ is assumed to be proportional to

\begin{equation}
\label{E-fmed}
f_{med} \sim \int_0^L d\zeta \epsilon^{3/4}(\zeta)F(\rho(\zeta), \alpha(\zeta))
\end{equation}
 where the integration runs over the eikonal path of the shower-initiating parton from production 
vertex to exit point from the medium. 

This scale-invariant modification manifestly preserves the self-similarity of the fragmentation function. In particular, from the point of view of the leading parton, this is a fractional energy loss mechanism since it is formulated as a function of the splitting variable $z$ only --- the average energy lost due to the medium effect is proportional to the initial energy. By construction, Eq.~(\ref{E-fmed}) makes the $L$ dependence of the medium effect approximately linear.

\begin{table}[htb]
\begin{center}
\begin{tabular}{|rccc|}
\hline
& YaJEM-DE & YaJEM-E & YaJEM+BW\\
\hline
self-similarity & broken & broken & preserved\\
$L$-dependence & non-linear & linear & linear\\
transverse shape & widened & narrowed & widened\\
soft gluon production & yes & no & yes\\
\hline
\end{tabular}
\end{center}
\caption{\label{T-summary}A summary of the essential physics characteristic of the scenarios used for the systematic study.}
\end{table}

A schematic summary of the physics characteristics of the three scenarios is given in Tab.~\ref{T-summary}. As apparent from the table, the choice represents quite distinct characteristics. Note again that the assumptions underlying YaJEM-E and YaJEM+BW can easily be criticized, however the purpose of their presence in this study is to illustrate which data sets will be sensitive to the characteristics displayed by them.

\section{Comparison with observables}

The basic strategy in the following comparison is that for all scenarios, the proportionality between $\epsilon^{3/4}$ and the relevant parameters ($\hat{q}, \hat{e}, f_{med}$) is chosen such that the CMS jet $R_{AA}$ measurement \cite{Roland} is reproduced at 100 GeV. Unless stated otherwise, no additional parameter is introduced or adjusted for the computation of any other observable.

\subsection{Jet observables}

\begin{figure}[htb]
\epsfig{file=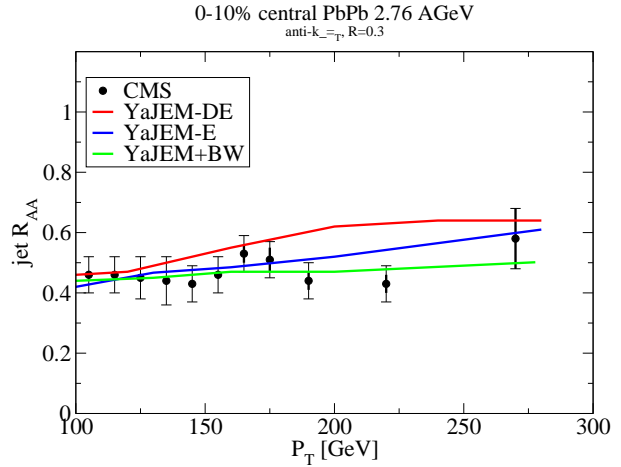,width=8cm}
\caption{\label{F-RAA-jet} Jet $R_{AA}$ for $R=0.3$ anti-$k_T$ jets computed in three different scenarios for parton-medium interaction compared with CMS data \cite{Roland}}
\end{figure}

A comparison of the three scenarios with the $P_T$ dependence of $R_{AA}$ for $R=0.3$ anti-$k_T$ jets is shown in Fig.~\ref{F-RAA-jet}.

As apparent from the figure, to first order the $P_T$ dependence is expected to be rather flat in all cases, with a hint of an increase in the case of YaJEM-DE. The differences between the scenarios are however of the order of the experimental errors, hence the data do not strongly discriminate the cases. Note that this finding is moderately surprising, as the different scenarios suppress jets with quite different mechanism --- while the main driving force in the case of YaJEM-DE and YaJEM+BW is the radiation of soft gluons which fall outside the $R=0.3$ cone, jets in YaJEM-E are actually collimated as no medium-induced gluons are radiated and energy from the shower is directly dissipated into the medium. Thus, flatness in $P_T$ is not necessarily expected.

 \begin{figure*}[htb]
\epsfig{file=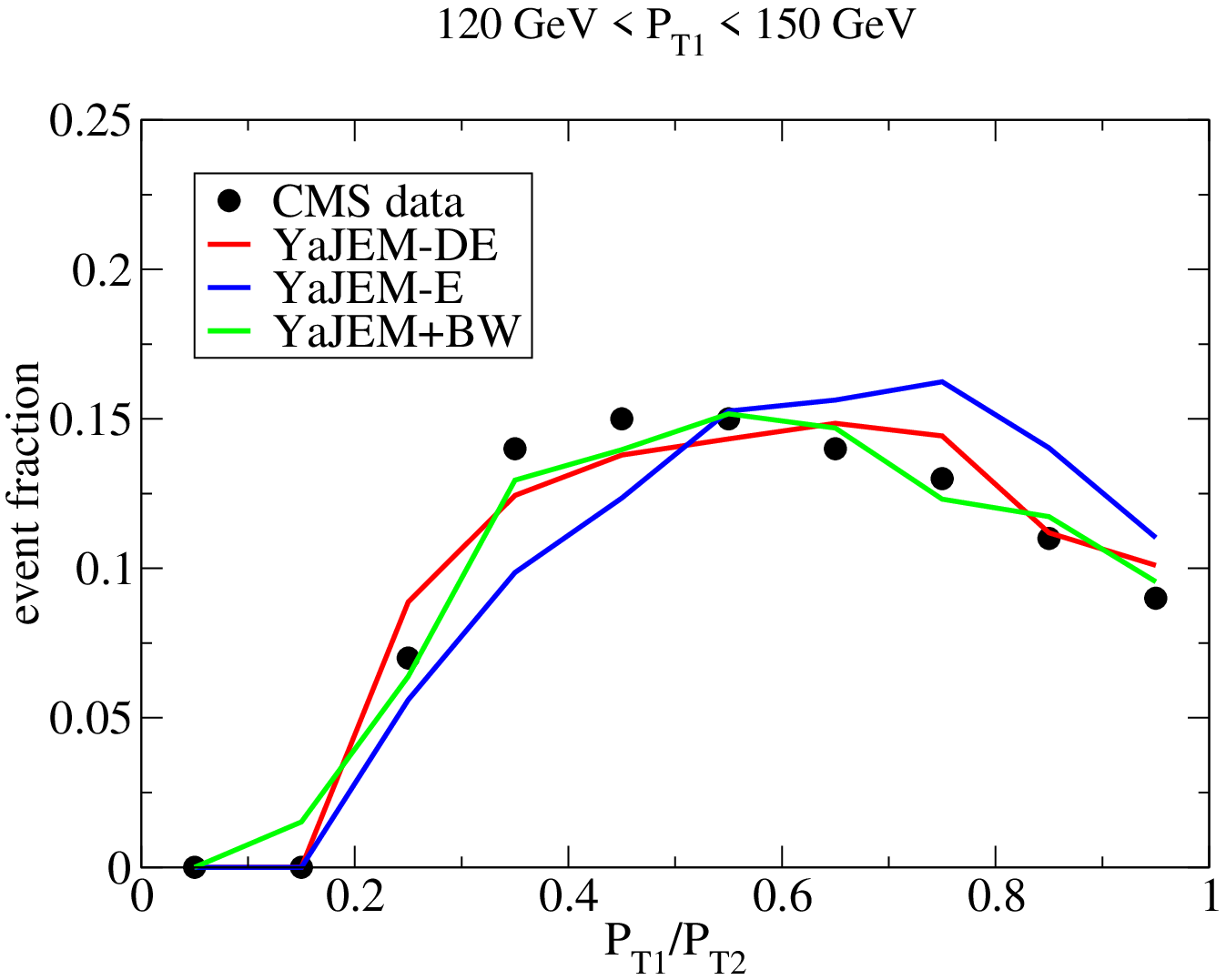,width=8cm}\epsfig{file=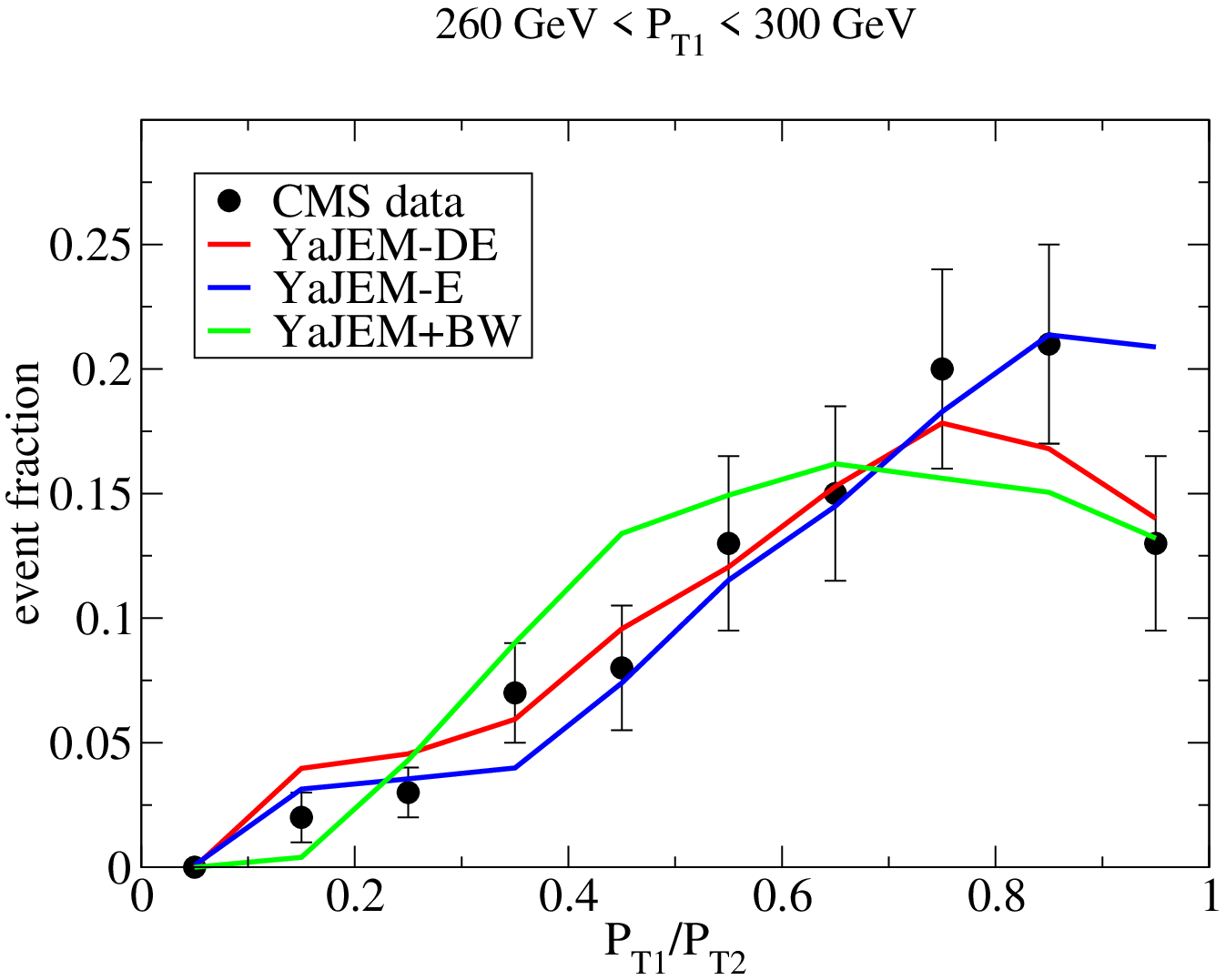,width=8cm}
\caption{\label{F-AJ} Dijet imbalance for $R=0.3$ anti-$k_T$ jets computed in three different scenarios for parton-medium interaction compared with CMS data \cite{CMS-jet-jet}}
\end{figure*}

A similar picture is apparent from studying jet-jet coincidences in terms of the dijet imbalance both for a trigger range of 120 to 150 GeV and for 260 to 300 GeV in Fig.~\ref{F-AJ}. While in the lower trigger energy range, the experimental statistics is good enough to create tension with the YaJEM-E scenario of the order of 20\%, overall the shape of the distributions is remarkably independent of the physics assumption on the quenching mechanism.

Note that the self-similar nature of YaJEM+BW is clearly apparent in the way the shape of the distribution in this scenario remains unchanged as a function of trigger energy. However, at the upper trigger energy range, the errors on the data are too large to decide whether a tension really exists or not. In all cases, the differences between the three scenarios are $O(20)$\%, which is surprisingly small given the substantially different physics assumptions underlying these scenarios, an observation which has previously been made in \cite{YaJEM-AJ}.

\subsection{Hadronic observables}

\label{Hadronic}

Let us now turn to purely hadronic observables. These differ in a factorized QCD picture from jet observables mainly in that they are sensitive to the fate of the leading parton only. This implies a dramatically different role of collinear splitting processes: While a collinear splitting of the leading parton leaves the jet energy invariant as the two daughters are clustered into the same jet, it reduces the energy of the leading parton. One might thus expect that hadrons are less robust against suppression by the medium than jets.

\begin{figure}[htb]
\epsfig{file=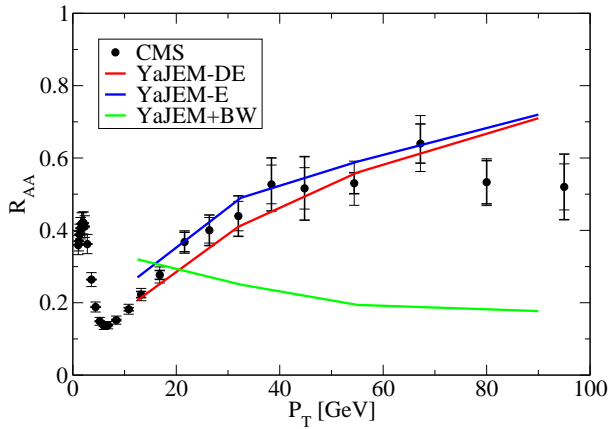,width=8cm}
\caption{\label{F-RAA-had} Charged hadron $R_{AA}$ in three different scenarios for parton-medium interaction compared with CMS data \cite{CMS-RAA}}
\end{figure}

This expectation is dramatically confirmed in Fig.~\ref{F-RAA-had} where the nuclear suppression factor of charged hadrons is compared with CMS data \cite{CMS-RAA}. Without explicit transfer of energy and momentum, the BW prescription results in what is in essence collinear emission of soft gluons from the leading parton, with the energy that is carried away being proportional to the original parton energy. Taken together, these two imply that BW can suppress hadrons more than jets, and that the suppression will grow stronger as hadron $P_T$ increases. This leads to a decreasing trend of $R_{AA}$ with charged hadron $P_T$ in striking disagreement with the data, demonstrating that hadron $R_{AA}$ is an excellent tool to rule out models based on collinear fractional energy loss.

Both other scenarios give a fair description of the data except at the highest $P_T$ (possible reasons for this are discussed in \cite{YaJEM-RAA}). This indicates that hadronic $R_{AA}$ is not in particular sensitive to the different pathlength dependence (a dependence on collimation vs. widening of the parton shower is not expected for an observable that looks at the leading shower parton only).

Observing more differentially, the hadron suppression can be studied as a function of the $v_2$ event plane of the bulk matter evolution and thus be expected to image the eccentricity deformation of the medium \cite{Abhijit-v2}. The suppression in-plane $R_{AA}^{in}$ and out-of-plane $R_{AA}^{out}$ can equivalently be cast into an angular averaged suppression and a harmonic coefficient $v_2$ (at high $P_T$), with the relation linking the descriptions being

\begin{equation}
\label{E-v2}
R_{AA}^{in} = R_{AA} (1+2v_2) \quad \text{and} \quad R_{AA}^{out} = R_{AA} (1-2v_2).
\end{equation}

As argued in \cite{Constraining} and tested systematically for a range of fluid dynamical models in \cite{SysHyd}, this $v_2$ is a sensitive probe of both pathlength dependence of the jet-medium interaction mechanism and the assumed geometry of the fluid dynamical medium. This strong dependence on the soft sector makes a direct extraction of pathlength dependence challenging. However, the $P_T$ dependence of $v_2$ can not be a property of the hydrodynamical medium but must be driven by the parton-medium interaction model only. In the following comparison, the normalization of $v_2$ is thus left open (in modeling, it can be varied substantially by changing the assumed equilibration time of matter, or introducing near $T_C$ enhancement of jet quenching as discussed in \cite{NTC}) and only the $P_T$ dependence is considered meaningful as a direct test of the parton-medium interaction.

\begin{figure}[htb]
\epsfig{file=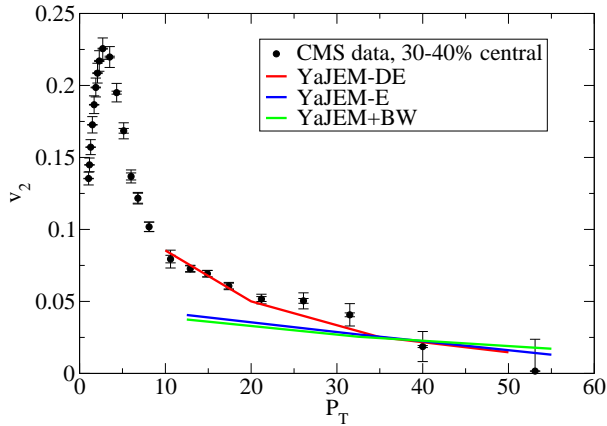,width=8cm}
\caption{\label{F-v2-had} Charged hadron$v_2$ in three different scenarios for parton-medium interaction compared with CMS data \cite{CMS-v2}}
\end{figure}

The result is shown in Fig.~\ref{F-v2-had}. As expected, the data discriminate clearly between the two models assuming linear pathlength dependence and YaJEM-DE with a non-linear dependence (note that below 10 GeV, $v_2$ is not necessarily driven by attenuation physics only but rather also  by hydrodynamical phenomena)

Unlike jet observables, hadron disappearance observables can hence be shown to yield consistently larger effect sizes $O(50\%)$ and even clear qualitative discrepancies with the trend of the data.

\subsection{Jet-hadron observables}

In order to retain sensitivity to possible collinear splittings and yet to be able to probe features of subleading shower partons, triggered correlation observables need to be studied. One possibility is to find jets in a certain energy range and then analyze the longitudinal (with respect to the jet axis) momentum distribution of hadrons which comprise the jet, which corresponds to the observables which are referred to by ATLAS and CMS as 'jet fragmentation function measurements' \cite{CMS-FF,ATLAS-FF}. We will consider this observable as an example in the following.

Note that such jet fragmentation function observables do not measure what a theorist usually would refer to as a fragmentation function, as there are several important differences: 1) In factorized QCD, the momentum fraction variable $z$ is defined with respect to the original parton energy, in the experimental analyses with respect to the observed jet energy (which may be substantially smaller and is on average different in vacuum and medium) 2) in theory, fragmentation functions are defined for quarks and gluons separately, whereas in the experimental analyses, a mixture of quark and gluon jets is measured, with this mixture being different in vacuum and medium and 3) fragmentation functions in theory are defined for an unbiased parton whereas the experimental analysis considers only the subset of jets falling into a certain energy range, and $R_{AA}^{jet} \sim 0.5$ indicates that the triggered set includes only half of the available jets. 

In the language of \cite{Bayesian}, the observable can therefore rather be seen as a near side intra-jet conditional yield ratio $I_{AA}$ given a triggered jet, and analoguously a correlation of away side (opposite to the trigger) hadrons with a jet can be studied \cite{STAR-jet-h}. 

\begin{figure}[htb]
\epsfig{file=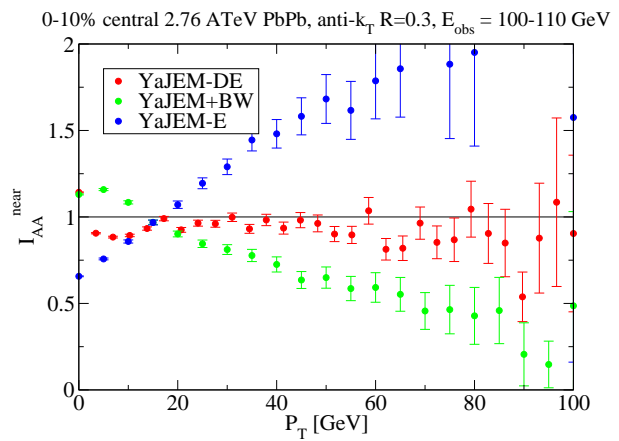,width=8cm}
\caption{\label{F-FF} Jet-triggered near side intra-jet hadron yield ratio as a function of hadron  momentum in three different scenarios for parton-medium interaction. }
\end{figure}

The result for the three models studied in this work is shown in Fig.~\ref{F-FF} (note that CMS and ATLAS data, each obtained for somewhat different kinematic cuts, lie to first approximation around unity).

Perhaps most striking is the fact that YaJEM-E leads to an \emph{enhancement} of the yield ratio across a large kinematic range. This is a clear indication that the observable is not actually the medium-modification of a fragmentation function, as using such a modification to compute hadron $R_{AA}$, enhancement above unity would inevitably follow, whereas the actual computation shows suppression (see Fig.~\ref{F-RAA-had}). The consistent interpretation is hence that 50\% of the jets are not seen (or rather, seen at significantly lower $P_T$ where they are subdominant) by the trigger, but that the triggered jet population has more hadron yield at higher momentum than vacuum jets inside an $R=0.3$ cone. This seemingly surprising outcome is hence an artifact of the shower collimation which happens in YaJEM-E (and goes away if the cone is widened to $R=0.7$).

YaJEM+BW shows quite the opposite trend, the fractional nature of energy loss model here suppresses the high $P_T$ yield much below what is expected in non-fractional energy loss models. Finally, the almost unmodified result found in YaJEM-DE results from a complicated cancellation of biases acting into different directions and is explained in  detail in \cite{Bayesian}.

Thus, while the observable clearly has no simple, intuitive interpretation, it is in fact just as sensitive $O(50\%)$ as purely hadronic observables to the different physics assumptions entering the models and probes both the narrowing vs. widening of showers and the non-fractional nature of energy loss.

\subsection{Hadron-hadron correlations at RHIC}

Let us now contrast the previous results which have all been obtained for LHC kinematics for 2.76 ATeV PbPb collisions with hadron-hadron back-to-back coincidence suppression as measured for 200 AGeV AuAu collisions at RHIC \cite{STAR-DzT}, an observable which has been identified as part of a set of key constraining observables for discriminating various scenarios of parton-medium interaction in \cite{Constraining}. In order to allow for a fair comparison, the relevant model parameters $(\hat{q}, \hat{e}, f_{med})$ are now re-adjusted to reproduce hadronic $R_{AA}$ at RHIC \cite{PHENIX-new} between 6 and 10 GeV. In principle, as demonstrated in \cite{hydro-LHC}, one could utilize the requirement of a description of data across different $\sqrt{s}$ to discriminate between models, however this incurs a sizable uncertainty in choosing the correct fluid dynamical bulk medium at both $\sqrt{s}$. For this reason, we do not pursue this idea here.

\begin{figure}[htb]
\epsfig{file=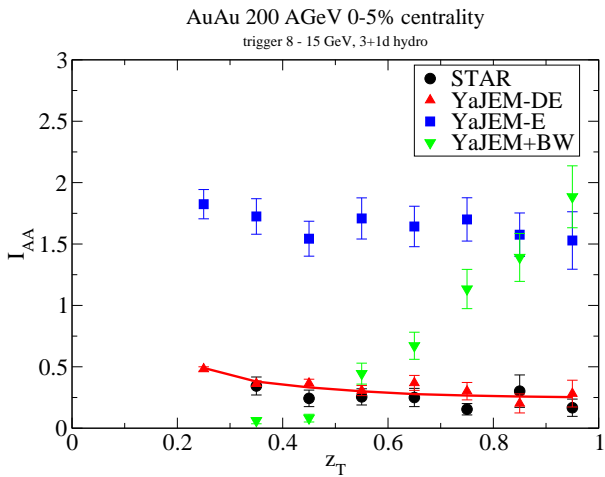,width=8cm}
\caption{\label{F-IAA} Hadron triggered away side hadron yield ratio as a function of associated hadron momentum over trigger momentum in three different scenarios for parton-medium interaction, compared with STAR data \cite{STAR-DzT}. }
\end{figure}

The resulting computation is shown in Fig.~\ref{F-IAA} where the medium over vacuum away side associate hadron yield ratio is shown as a function of $z_T= P_{assoc}/P_{trigger}$. Both YaJEM+BW and YaJEM-E show dramatic differences (of a factor five and more) to the data, with in addition YaJEM+BW exhibiting the completely wrong trend. As might be expected from a key observable, there is hence strong discriminating power between different models.

As discussed in some length in \cite{Bayesian}, the main reason for the strong discriminating power is a coincidence of biases which enhances the sensitivity to  three major physics ingredients of the parton-medium interaction model --- kinematic shift, pathlength dependence and quark vs. gluon energy loss.

Let us start with the pathlength dependence. Since the trigger for this analysis is a hadron, the analysis is biased towards showers in which the near side leading parton has not lost much energy. For a steeply falling parton spectrum as relevant at RHIC kinematics, even a small energy shift translates into a strong yield suppression at given $p_T$, this is progressively less true for the harder spectra at LHC kinematics. For a medium modification which grows with medium density and with in-medium pathlength, the implication is that the trigger condition will preferentially select events in which the hard vertex was close to the surface and the trigger parton moved only a short path outward through the medium. This is known as 'surface bias'. The surface bias is much stronger for a non-linear pathlength dependence of the medium modification (as exhibited by YaJEM-DE) than for a linear one. However, placing the production vertex near to a surface implies that the majority of away side partons is forced onto a long path through the medium center, which in the non-linear case also receives extra weight. For this reason, h-h correlations at RHIC energies are maximally sensitive to the pathlength dependence of models and generically all models based on linear pathlength dependence overshoot the data by a large margin, as already realized in \cite{ElasticPhenomenology}. The surface bias as a sensitive probe of the pathlength dependence accounts for the fact that both YaJEM-E and YaJEM+BW on average overshoot the data.

The second relevant effect is the medium-induced shift in the relation between trigger $P_T$ and hard event parton $p_T$. At RHIC, the leading hadron typically contains the fraction $z\sim0.7$ of the leading parton energy in the vacuum and $z\sim 0.5$ in the medium \cite{Dihadron}. Within the limited kinematic range probed by the observable, these relations do not change substantially for YaJEM-DE and YaJEM-E where the fragmentation function is not self-similar and the strength of the medium effect is primarily set by medium effects, but they do vary strongly in YaJEM+BW where the amount of energy lost from the leading parton is approximately proportional to its energy. It is this variation which changes the shape of  $I_{AA}(z_T)$, just as it altered the shape of the hadronic  $R_{AA}(P_T)$ as shown in Fig.~\ref{F-RAA-had}.

The third major effect, also discussed in \cite{Dihadron}, is the observation that for the trigger kinematics at RHIC, the away side parton type is biased to be a gluon. Due to the softer fragmentation of gluons in general and the stronger coupling of gluons to the medium, this leads to extra suppression, causing a very low value of $I_{AA}$. However, since all three scenarios investigated here take the difference between quarks and gluons into account consistently, this is not a discriminating factor relevant for the interpretation of  Fig.~\ref{F-IAA}.

Note that the observable is in addition at low $z_T$ sensitive to the presence or absence of subleading shower partons --- as demonstrated in \cite{Dihadron}, pure leading parton energy loss models miss the upturn of $I_{AA}$ below $z\sim 0.4$ and instead result in a downturn.

In summary, this example shows that there is a subset of key observables in which biases selectively magnify the relevant differences between models and give a discriminating power between models which is an order of magnitude larger than with other observables which are less sensitive to model assumptions.

\section{Discussion}

\subsection{Other observables}

There are many more high $P_T$ observables available which have not been included into the series of comparisons presented above, chiefly because their properties (in the language of \cite{Bayesian}, their bias structure) are known, and hence it can be at least qualitatively inferred what properties of parton-medium interaction they are (in-)sensitive to.

Consider for instance the centrality dependence of angular-averaged jet or hadron $R_{AA}$. Given that the hydrodynamical medium is constrained by the variations in bulk spectra and $v_2$ independently of the hard physics, it would seem that testing the centrality dependence of the suppression factor poses non-trivial constraints to models. However, one should consider that both endpoints of the centrality evolution are already constrained --- the situation for the most central collisions is shown in Figs.~\ref{F-RAA-jet}, \ref{F-RAA-had}, whereas for very peripheral collisions eventually only a single pair of nucleons collides, at which point final state interaction must cease and $R_{AA} \approx 1$ can be inferred. The centrality dependence is hence just an interpolation between these points, and while there are interesting structures in the data in detail, it is hard to see how a model in which the medium modifications in some way scale with pathlength and density would be able to miss this interpolation by more than 20\%. As evidence supporting this assertion, \cite{Djordjevic} may be taken where the centrality dependence of angular averaged hadronic $R_{AA}$ can be reproduced at RHIC and LHC just based on density scaling of the medium, i.e. without a genuine fluid dynamical computation including transverse expansion of the medium.

In an effort comparable to the one presented here, in \cite{Hybrid}, three parametric models of parton energy loss (strong coupling, radiative, collisional) have been tested against a variety of jet observables, including the centrality dependence of jet $R_{AA}$ and the dijet imbalance. The results obtained are consistent with the results presented here, the only jet observable with discriminating power found is the jet-triggered near side intra-jet $I_{AA}$.

Let us then briefly turn to other classes of observables. As discussed in \cite{Bayesian}, the bias structure of jet-h correlations can be made similar to that of hadron-hadron correlation by choosing a suitable jet definition, i.e. a discriminating power equivalent to Fig.~\ref{F-IAA} could be expected for this observable, whereas $\gamma$-hadron correlations do not suffer from a surface bias effect and would hence chiefly discriminate YaJEM-BW based on the fractional energy loss, while YaJEM-DE and YaJEM-E would be distinct only on the amount of subleading radiation produced at low $z_T$, making photon-triggered correlations somewhat less discriminative.

Instead of considering the longitudinal structure of a found jet, also the transverse structure might be investigated, the CMS collaboration has done this in terms of a jet shape measurement \cite{CMS-jetshape}. However, as \cite{YBW} demonstrated, at least YaJEM+BW is quite compatible with the measurement once the biases due to jet finding are taken into account properly.

\subsection{Theoretical uncertainties}

Any discussion on the discriminating power of various observables is incomplete without a statement at what level of accuracy various models need to be separated. This is best apparent from observables like the dijet imbalance at low $P_T$ (Fig.~\ref{F-AJ}, left panel) where experimental errors are vanishingly small. Assuming vanishing theoretical uncertainty, this plot would in fact rule out all scenarios except YaJEM+BW with high confidence. However, theoretical uncertainties do not vanish, they are merely hard to assess.

An overview of what is known about the various uncertainties is given in \cite{Constraining}, here just a few essential points are summarized. One source of uncertainties is of numerical nature, these arise from finite sampling size in Monte-Carlo strategies and finite resolution in numerical integrations. Where such uncertainties are an issue, they have been shown as theoretical error bars in the plots.

A second class of uncertainties has to do with the  pQCD computation underlying the initial state of any parton-medium interaction modeling and in particular with scale choices for the parton distribution functions (PDFs) as well as correct error propagation of the PDF and nuclear PDF errors. Typically such pQCD uncertainties are relatively small given the nature of observables as ratios and conditional probabilities, observable-dependent between 5 and 10\%.

The next class of errors has to do with the validity of assumptions frequently made in modelling final state parton-medium interaction. An example for that is eikonal parton propagation, which corresponds to a small uncertainty $<5$\%.

Uncertainties with regard to the modeling of the medium are conceptually somewhat difficult. In tomographic measurements, rather than being an uncertainty, the response to the assumed background medium is the objective of the measurement, whereas in measurements aimed to constrain parton-medium interaction, the same response is a systematic uncertainty. As alluded to in section \ref{Hadronic} and observed in \cite{Constraining,SysHyd}, in principle a sizable uncertainty, approaching 100\% for some observables like the magnitude of hadronic $v_2$, is associated with the choice of the background medium in which the parton-medium interaction model is tested. This uncertainty is  smaller for other observables, for instance for hadron-hadron coincidence $I_{AA}$ it is known to be about 20\% \cite{Dihadron} and for jet coincidences less than 5\% \cite{YaJEM-AJ}. However, utilizing the tighter constraints which are posed to modern fluid dynamics computations by observables such as the event-by-event fluctuations of bulk matter $v_2$ \cite{v2} may reduce this uncertainty below what constraints from demanding agreement to bulk spectra and event-averaged $v_2$ yield. In a similar way, possible near-$T_C$ enhancement of jet quenching \cite{Liao,Liao2,Liao3}, if not the objective of the measurement, introduces a 30\% uncertainty on the magnitude of $v_2$ and a $<5\%$ uncertainty on the mean $R_{AA}$.

Finally, the decision to fit the parameters $(\hat{q}, \hat{e}, f_{med})$ to some region of experimental data to determine the overall strength of jet quenching introduces a systematic uncertainty of the magnitude of the combined statistical and systematical experimental errors of the data.

However, note that all these theoretical uncertainties are (anti-)correlated across different observables in complicated ways. To give a simple example for such an anticorrelation, changing the spectral slope of the primary QCD spectrum within the uncertainties might lead to a decrease in $R_{AA}$, but due to the resulting change in the kinematic relation between trigger hadron and parent parton cause a kinematic bias which increases $I_{AA}$ in dihadron correlations.

In summary, while a general assessment of the theoretical uncertainties in a systematic way is an extremely challenging problem and has never been attempted so far, the order of magnitude of the uncertainty for the uncertainty of the observables discussed in this work can be estimated to be at least 10-20\%. The implication is that a 20\% effects are not sufficient to solidly discriminate between models and rule out certain scenarios of parton-medium interaction, as tweaking of parameters within plausible ranges might move the theoretical curves within that range. 50\% effects are clearly enough to distinguish between the gross differences exemplified by the scenarios studied here, but to distinguish more subtle differences between models, the large factors of selectively biased key observables are the only viable avenue. 

\section{Conclusions}

The study of several types of high $P_T$ observables for three rather different physics scenarios of parton-medium interaction presented in this work illustrates a number of points.

First, not every observable is sensitive to the same physics properties. For instance while charged hadron $R_{AA}(P_T)$ distinguishes models cleanly along the lines of fractional vs. fixed scale modification, $v_2(P_T)$ keys primarily on linear vs. non-linear pathlength dependence. This is in principle a desirable property which helps selectively probing relevant physics, however argues caution in making conclusions about model validity based on a subset of data. In an extreme case, the fact that a model agrees with the whole set of jet $R_{AA}$ and dijet imbalance measurements in addition to the intra-jet correlated hadron yield and hadronic $R_{AA}$ does still not imply that this model has the correct pathlength dependence, as none of these observables are sensitive to pathlength dependence but rather probe other properties. Similarly, demonstrating good agreement between model and jet $R_{AA}$ and imbalance even across a significant centrality range may still not imply strong constraints for the model.

Second, observables focusing on final state hadron yield are more sensitive than observables focusing on jet yield. This may seem counter-intuitive, as the jet yield involves more of the shower particles, however it can be understood by considering the different sensitivity to near-collinear splittings. Another way of expressing this is that jet definitions are designed to suppress physics around $\Lambda_{QCD}$, i.e. soft gluon emission or hadronization and instead be comparable to hard QCD calculations on the parton level. As a result, clustering systematically also suppresses the physics of parton-medium interaction modifying the QCD shower evolution at a scale of the temperature $T \sim \Lambda_{QCD}$. 

Third, triggered correlations where a final state hadron is observed provide the highest discriminating power of all observables, however are usually counter-intuitive to interpret. The reason for both is the same --- the nature of triggered correlations as conditional probabilities implies a bias structure that can be utilized to selectively magnify certain aspects of the physics.

Finally, there are selective advantages to measurements both at RHIC and LHC kinematics. While at RHIC, the steeply falling spectrum implies that biases act much stronger in triggered correlations (e.g. in terms of surface bias \cite{Bayesian}), the large kinematic range accessible at the LHC allows to observe different scaling behaviour with the initial parton energy (for instance fractional energy loss) cleanly. Only the interplay between low and high $\sqrt{s}$ supplies a meaningful set of key observables constraining essentially all aspects of the interaction between parton showers and the medium. 

Based on these results, a suitable strategy suggesting itself for a most constraining set of measurements is hence to use clustering to access the properties of the primary hard event or to provide a selectively biased trigger, and then focus on hadrons correlated with the trigger object to study the softer physics of parton-medium interaction and energy dissipation into the medium.

\begin{acknowledgments}
  This work is supported by the Academy researcher program of the
Academy of Finland, Project No. 130472. 
 
\end{acknowledgments}

\end{document}